\documentclass[conference]{IEEEtran}
\IEEEoverridecommandlockouts
\usepackage{cite}

\usepackage{amsmath}
\usepackage{amsfonts}
\usepackage{multirow}
\usepackage{amssymb}
\usepackage{graphicx}
\usepackage{epstopdf}
\usepackage{enumerate}
\usepackage{textcomp}
\usepackage{xcolor}
\usepackage{color}

\usepackage{multirow}

\usepackage{algorithm}
\usepackage{algpseudocode}
\usepackage{mathtools, cuted}
\usepackage{lipsum}

\def\E#1{\ensuremath{\text{E}\left \{#1 \right \}}}

\newcommand{\frm}[1]{\langle #1\rangle}

\def\BibTeX{{\rm B\kern-.05em{\sc i\kern-.025em b}\kern-.08em
    T\kern-.1667em\lower.7ex\hbox{E}\kern-.125emX}}

\newcommand{\tof}{\delta}
\newcommand{\per}{\Delta}
\newcommand{\ti}{\tau}
\newcommand{\ts}{\eta}
\newcommand{\an}{\mathbf{a}}
\newcommand{\p}{\mathbf{p}}
\newcommand{\ve}{\mathbf{v}}
\newcommand{\m}[1]{{{#1}^\star}}

\begin{document}

\title{Scale up to infinity: the UWB Indoor Global Positioning System
\\[-2.5ex]
}

\author{
	\IEEEauthorblockN{Luca Santoro, Matteo Nardello, Davide Brunelli, Daniele Fontanelli}
	\IEEEauthorblockA{\textit{Department of Industrial Engineering}, 
		\textit{University of Trento}, Trento, Italy \\
		name.surname@unitn.it
}
%\author{\IEEEauthorblockN{1\textsuperscript{st} Luca Santoro}
%\IEEEauthorblockA{\textit{Dept. of Industrial Engineering} \\
%\textit{University of Trento}\\
%Trento, Italy \\
%luca.santoro@unitn.it}
%\and
%\IEEEauthorblockN{2\textsuperscript{nd} Matteo Nardello}
%\IEEEauthorblockA{\textit{Dept. of Industrial Engineering} \\
%\textit{University of Trento}\\
%Trento, Italy \\
%matteo.nardello@unitn.it}
%\and
%\IEEEauthorblockN{3\textsuperscript{rd} Davide Brunelli}
%\IEEEauthorblockA{\textit{Dept. of Industrial Engineering} \\
%\textit{University of Trento}\\
%Trento, Italy \\
%davide.brunelli@unitn.it}
%\and
%\IEEEauthorblockN{4\textsuperscript{th} Daniele Fontanelli}
%\IEEEauthorblockA{\textit{Dept. of Industrial Engineering} \\
%\textit{University of Trento}\\
%Trento, Italy \\
%daniele.fontanelli@unitn.it}
% \and
% \and
% \and
\\[-2.5ex]
\thanks{This article has been accepted for the 2021 IEEE International Symposium on Robotic and Sensors Environments (ROSE). Citation information: L. Santoro, M. Nardello, D. Brunelli and D. Fontanelli, "Scale up to infinity: the UWB Indoor Global Positioning System," 2021 IEEE International Symposium on Robotic and Sensors Environments (ROSE), 2021, pp. 1-8, doi: 10.1109/ROSE52750.2021.9611770. }
}

\maketitle

\begin{abstract}
  Determining assets position with high accuracy and scalability is
  one of the most investigated technology on the market.  The
  accuracy provided by satellites-based positioning systems (i.e.,
  GLONASS or Galileo) is not always sufficient when a decimeter-level
  accuracy is required or when there is the need of localising
  entities that operate inside indoor environments. Scalability is
  also a recurrent problem when dealing with indoor positioning
  systems. This paper presents an innovative UWB Indoor GPS-Like local
  positioning system able to tracks any number of assets without
  decreasing measurements update rate. To increase the system's accuracy
  the mathematical model and the sources of uncertainties are investigated.
  Results highlight how the proposed implementation provides 
  positioning information with an absolute maximum error
  below 20 cm. Scalability is also resolved thanks to DTDoA
  transmission mechanisms not requiring an active role from the asset
  to be tracked.
\end{abstract}

\vspace{1em}


\section{Introduction}
\label{sec:Intro}

As an accurate positioning system is the mainstay of most mobile 
robot applications~\cite{SiegwartBook}, researchers have proposed 
to implement different LPS to provide an indoor localisation 
infrastructure. As described in~\cite{PositioningSurvey} positioning 
systems can be classified into two macro classes: \textbf{1)} Global 
Position Systems (GPS) and \textbf{2)} Local Positioning Systems (LPS). 
In the last decades, significant progress was registered in the 
development of positioning systems~\cite{karl2007celestial}, 
especially to find a low-cost, accurate and local alternative to GPS 
for the indoor environment~\cite{6153154, 8950421,app11010279,s21031002}. 
This also thanks to the ground-breaking improvements made to industries 
such as healthcare~\cite{YIN20163, 6720693}, livestock 
farming~\cite{1550147720921776, 8074146}, warehousing~\cite{8053982, 9330058} 
and, more recently, in the robotic field~\cite{Obeidat2021, 8603685}.\newline
LPSs can be categorised~\cite{PositioningSurvey} by: \textit{1) 
Computation approach}: centralised, decentralised; \textit{2) 
Environment type}: underwater, outdoor, and indoor; \textit{3) 
Communication Technology}: optical, acoustic and radio frequency. 
For example, in~\cite{AcousticTril} authors localise the position 
of an acoustic source in underwater environments. In~\cite{IRKalman} 
authors improve the navigation system providing more accurate pose 
information to aid the odometry using infrared sensors. Based on 
optical waves as transmission medium, in~\cite{VLL} conventional 
lights are used to enable the localisation. 

A promising approach to localise a moving entity -- with decimetre
accuracy level -- is assessing Radio Frequency (RF) signal
propagation. Different RF positioning solutions have been tested; like
RFID~\cite{RFIDPos}, WLAN~\cite{WLANPos} and cellular
network~\cite{CellularPos}. In the class of RF technology,
ultrawide-band (UWB) has attracted increasing interest due to its
excellent characteristics, like robustness to multipath error,
obstacle penetration, high accuracy and low cost~\cite{s16050707}. In
the case of RF-based localisation systems the fundamentals ranging
techniques to estimate the distance from two nodes are:
\textit{Received Signal Strength} (RSS); \textit{Time of Arrival}
(ToA); \textit{Time Difference of Arrival} (TDoA) and \textit{Angle of
  Arrival} (AoA). RSS is the simplest approach to estimate the
distance between two nodes~\cite{RSSI,nardello-covid}. In this case, the path loss
model is exploited to estimate the distance by measuring the signal's
received power. ToA implementations are based on the time of arrival
of a signal, which allows to determine the distance from 
the RF wave propagation speed. The location information
can be determined using algorithms like trilateration or
multilateration solving the associated geometrical problem (i.e.,
circle or sphere intersections)~\cite{TriProb}. In TDoA, as for the
ToA case, the distance estimation is based on the propagation
speed. However, in this case, the difference in receiving times of
a signal is considered, leading to a geometrical problem based on
hyperbolas~\cite{TDOAProb}. AoA implementation exploits the
capabilities of unknown nodes of detecting the angles of incoming
signals. One common approach to obtain AoA measurements is to use an
antenna array on each sensor node. It is then possible to discover
both the position and the orientation by exploiting the angle
measurements~\cite{4068140}.

In the UWB localisation systems, ToA and TDoA ranging techniques are
commonly adopted. In the case of the ToA approach, two different
communication schemes can be implemented: Single-side (SS) or
double-side (DS) two way ranging (TWR)~\cite{SingleDoubleTWR}. With
TWR, a synchronisation mechanism between nodes is not required. In
this case, an accurate calibration of the crystal oscillators is
sufficient to increase the measurement accuracy, especially in
SS-TWR. The drawback of this approach is the maximum achievable 
measurements data rate~\cite{MagnagoCPPF19iros} that depends on the 
number of messages that the tag has to exchange with the anchors. For example, 
DS-TWR requires four messages to be exchanged between each tag and 
each. This to obtain just a single ranging information. It is thus clear that 
the higher the number of the tag, the lower the update rate.

\begin{figure}[t]
	\centering
	\begin{tabular}{cc}
		\includegraphics[width=0.45\columnwidth]{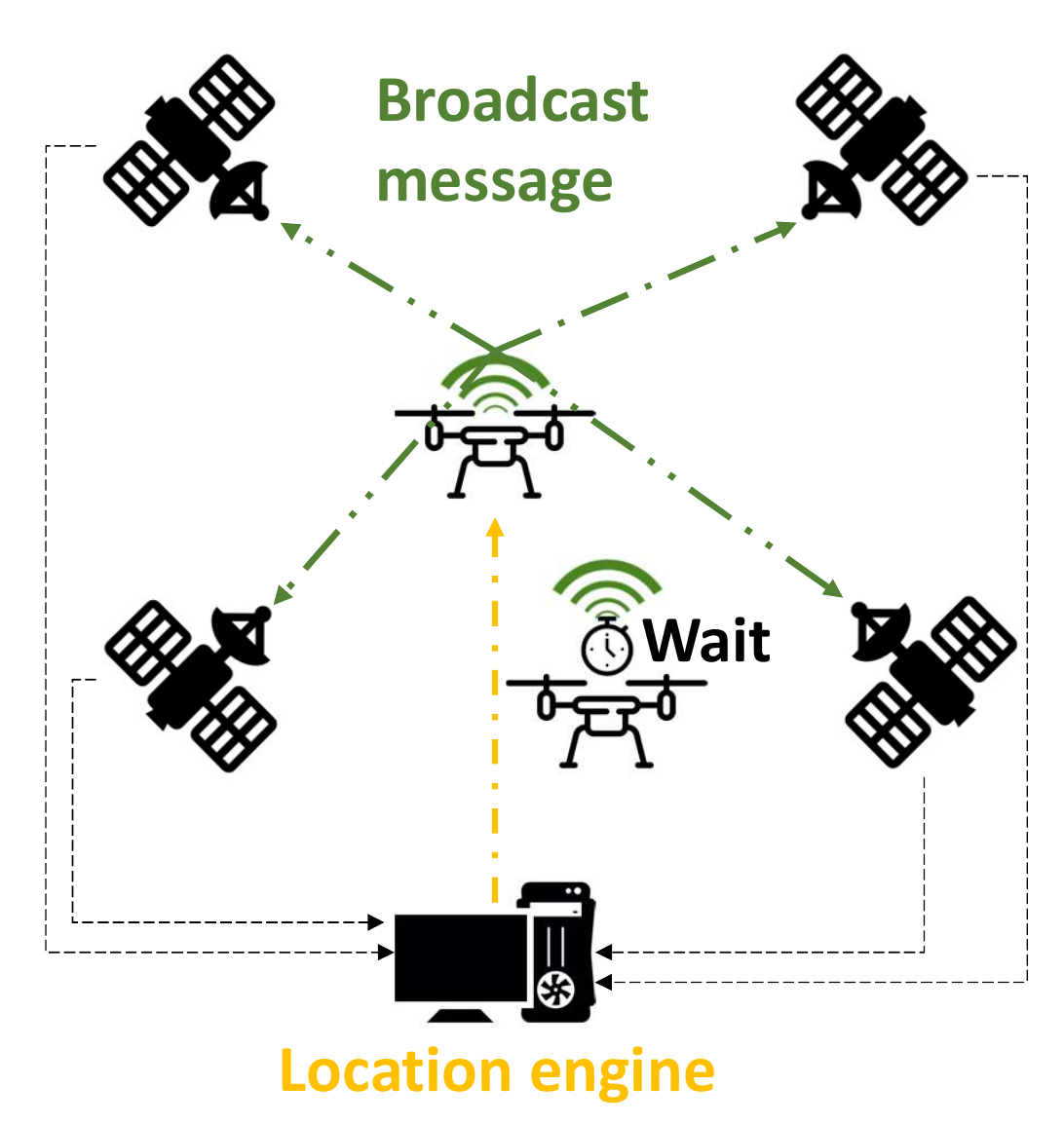} &
		\includegraphics[width=0.45\columnwidth]{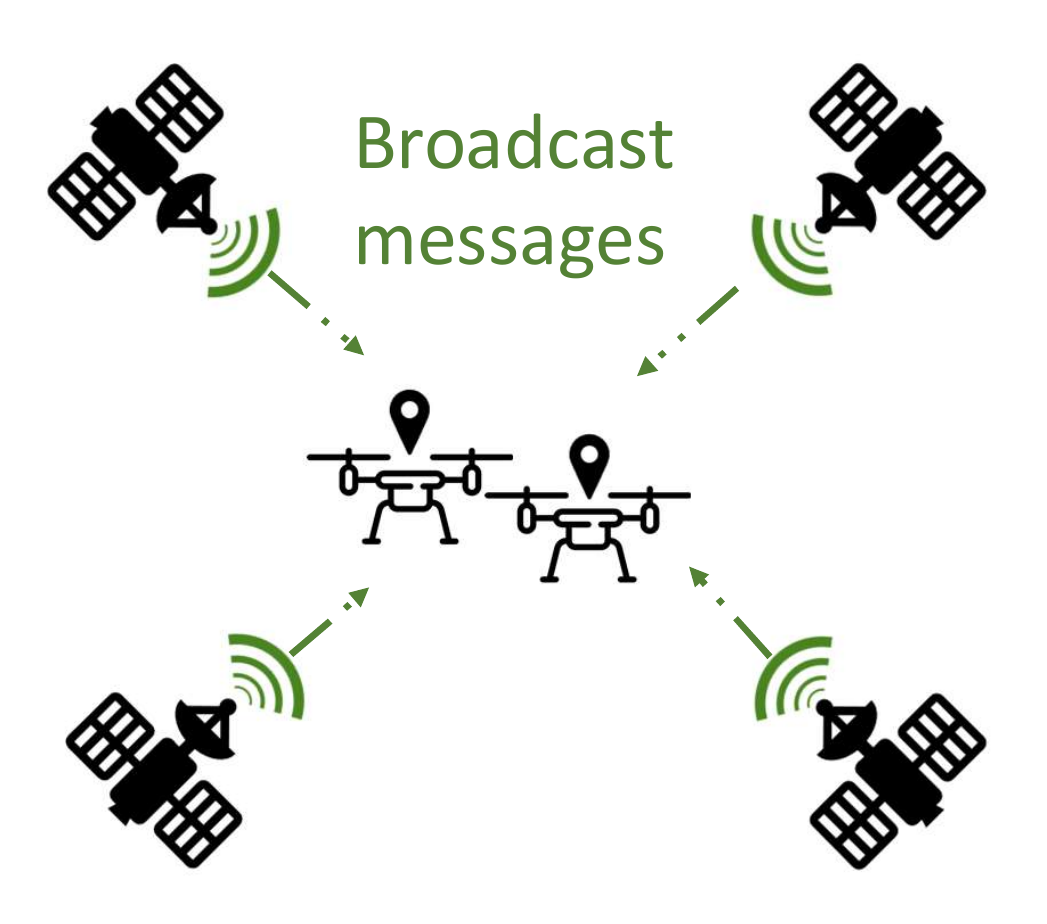}
		\\
		(a) & (b)
	\end{tabular}
	\caption{In (a) the traditional UTDoA approach where the tag emits the broadcast message, thus limiting the scalability of the system. In (b) the implemented DTDoA approach, where the anchors emits the messages used by the entity to localisation awareness.}
	\label{fig:Scenario}
	\vspace{-1em}
\end{figure}

To overcome this issue, the TDoA approach is investigated. In this
case, crystal oscillator trimming is not enough to achieve the desired
accuracy. A tight synchronisation between nodes is required. As for
the ToA approach, also for TDoA two possible schemes can be
implemented. A centralised, or uplink, TDoA (UTDoA) and a
decentralised, or downlink, TDoA (DTDoA). In the Figure~\ref{fig:Scenario} is depicted the DToA and UTDoA approaches. In the first approach a tag
emits an UWB signal and the difference in the reception times at 
anchors side is used to calculate the position of the tag with respect
to a reference point~\cite{UTDoA}. With this method, the update rate
of pose information is mainly determined by the number of entities to 
be tracked and the entity's blinking duty cycle. The drawbacks of this
approach are mainly two. The first disadvantage is that the position 
information resides at the infrastructure side (i.e., the anchors know
tags positions, but tags do not), requiring a phase where this 
information must be shared down to the tag side. The second 
disadvantage is that the number of tags affects the achievable 
update rate. The higher the number of the tags, the lower the 
achievable update rate. To overcome these two drawbacks, researchers 
have proposed the DTDoA approach. In this case, the system 
mimics a common GPS system, where  the anchors -- as the satellites -- 
continuously broadcast time-stamped messages that can be received 
by listening tags/robots. In~\cite{Chorus,8250062} 
authors developed a DTDoA system, where a tag can determine 
its position with respect to a reference point 
exploiting the concurrent ranging (i.e., anchors simultaneously emit 
a UWB signal). Unfortunately, due to hardware limitations and the 
precision of the timestamps, the system can achieve a maximum 
position accuracy in the order of a couple of meters. To mitigate 
this problem in~\cite{STDoA} the authors exploit the idea that each 
anchor sequentially blinks a message, reducing the error on the 
estimated position below 1 meter.
%-%
%-%
This paper presents innovative DTDoA ranging techniques solving both 
accuracy and scalability problems of the current state of the art.
Results highlight how the systems can theoretically scale to infinity
(i.e., any number of assets can be tracked), improving the 
measurement accuracy with an error in the range of 20cm, at worst. 

\noindent The main contributions of this paper are:

\begin{itemize}
	\item
		The mathematical analysis of the transmission model and the characterisation of the source of uncertainties to improve the system accuracy;
	\item
		The validation of the mathematical model exploiting low-cost COTS UWB radios.
\end{itemize} 

The rest of the paper is organised as follows. Section~\ref{sec:models} 
present and discuss the mathematical model of the implemented 
DTDoA ranging scheme. In Section~\ref{sec:uncertainties} system 
uncertainties and sources of error are discussed and analysed. 
Experimental results and evaluation are presented in 
Section~\ref{sec:ExperimentalResults}. Section~\ref{sec:conclusion} 
closes this work with final remarks and possible future improvements.

\section{Models}
\label{sec:models}

In this section we are going to present the mathematical model of the
proposed LPS.  To this end, we consider an environment with a master
UWB, a set of $n$ anchors $\an_i$ and a tag. We can thus denote $t$ as
the actual, ideal time and with $\ti$ the time measurement from either
the master $\ti^m(t)$; the $i$-th anchor $\an_i$ as $\ti^i(t)$; or the
tag $\ti(t)$. Since we do not have an external time reference, we can
assume that the time measurements of the master are the reference
signal for the UWB positioning algorithm, following the simplified
clock model presented in~\cite{FontanelliMWOS11}, i.e.
\begin{equation}
  \label{eq:TauMaster}
  \ti^m(t) = o_m + \nu_m t ,
\end{equation}
where $o_m$ is the time offset of the master and
$\nu_m = \frac{\overline{f}_m(t)}{f_m}$ the normalised clock rate with
respect to the ideal time (i.e., the ratio between the instantaneous
frequency of the local oscillator $\overline{f}_m(t)$ and the
corresponding nominal value $f_m$, usually on the order of some part
per million (ppm)~\cite{FontanelliMWOS11}). Similarly, for the $i$-th
anchor $\an_i$ we have
\begin{equation}
  \label{eq:TauAnchor}
  \ti^i(t) = o_i + \nu_i t ,
\end{equation}
where $o_i$ and $\nu_i$ are the $i$-th offset and clock rate of the
time of the anchors with respect to the ideal time. Finally, for the
tag we have
\begin{equation}
  \label{eq:TauTag}
  \ti(t) = o + \nu t .
\end{equation}

\subsection{Anchors clock analysis}

The main idea underlying this approach is that no message exchange
should be carried out from the tag to the anchors, but only from the
anchors to the tag. This to ensure infinite scalability in terms of trackable 
number of tags. In the ideal case, the quantities $o_i$, $\nu_i$ (with
$i = 1,\dots,n$) in~\eqref{eq:TauAnchor} with respect to the master
reference time are retrieved through the following synchronisation
algorithm: starting at a generic time $\overline t$, the master anchor sends
two messages $\ti^m(\overline t)$ and
$\ti^m(\overline t + \per_{i,m})$ to $\an_i$, whose timestamps at the
receiving side are $\ti^i(\overline t + \tof_{i,m})$ and
$\ti^i(\overline t + \per_{i,m} + \tof_{i,m})$, where $\tof_{i,m}$ is
the Time of Flight (ToF) from the master to the anchor. By denoting
with $[x_m,y_m]^T$ and $\an_i = [x_i,y_i]^T$ respectively the master
and the anchor known Cartesian coordinates in the $X_w\times Y_w$
plane with respect to a fixed reference frame
$\frm{W} = \{X_w, Y_w, Z_w\}$, it turns out that
\[
  \rho_{i,m} = \| [x_m,y_m]^T - \an_i \| = \sqrt{(x_m - x_i)^2 + (y_m
    - y_i)^2} ,
\]
is the distance among the two anchors (where $\|\cdot\|$ is the usual
Euclidean norm). Therefore, assuming that $c$ is the known propagation
speed of the radio frequency signal in LOS conditions, we have that
$\tof_{i,m}$ can be obtained as
\begin{equation}
  \label{eq:tof}
  \tof_{i,m} = \frac{\rho_{i,m}}{c} .
\end{equation}
As a consequence, using~\eqref{eq:TauMaster} and~\eqref{eq:TauAnchor}
we can derive the relative clock rate
\begin{equation}
  \label{eq:Synchro}
  \overline \nu_{i,m} = \frac{\nu_i}{\nu_m} = \frac{\ti^i(\overline t + \per_{i,m} + \tof_{i,m}) -
    \ti^i(\overline t + \tof_{i,m})}{\ti^m(\overline t + \per_{i,m})
    - \ti^m(\overline t)},
\end{equation}
and the relative offset
\begin{equation}
  \label{eq:SynchroOffset}
  \begin{aligned}
    & \ti^i(\overline t + \tof_{i,m}) - \overline \nu_{i,m} \ti^m(\overline t) - \tof_{i,m} = \\
    & = o_i - \overline \nu_{i,m} o_m -
    (1 - \nu_i) \tof_{i,m} = \overline o_{i,m} = \\
    & = \ti^i(\overline t + \per_{i,m} + \tof_{i,m}) -
    \overline \nu_{i,m} \ti^m(\overline t + \per_{i,m}) - \tof_{i,m} . \\
  \end{aligned}
\end{equation}
By assuming that the clock rates $\nu_m$ and $\nu_i$
are approximately constant among two synchronisation periods (usually
executed every tens of seconds), and that the master and the anchors
do not change their relative positions (i.e., the ToF $\tof_{i,m}$ is
constant), we can derive that the relative offset $\overline o_{i,m}$ 
is constant as well. Furthermore, the term $(1 - \nu_i) \tof_{i,m}$ is usually
negligible with respect to the relative offset between the master and
the anchor. In turns, by using $\overline \nu_{i,m}$ -- representing the
correction factor that transforms the timestamped measured quantities
in master time-scale into the $i$-th anchor time-scale -- a generic
timestamp $\ti^i(t)$ of an anchor can be translated to the master
time-scale using~\eqref{eq:TauMaster} and
\begin{equation}
  \label{eq:AnchorSync}
  \frac{\ti^i(t) - \overline o_{i,m}}{\overline \nu_{i,m}} = o_m + \nu_m t
  + \frac{1 - \nu_i}{\overline \nu_{i,m}} \tof_{i,m} = \ti^m(t) +
  e_{i,m}.
\end{equation}

\subsection{Tags clock analysis} 
Due to the DTDoA approach used, also tag's clock have to be corrected.
Let's consider the tag at time $t$ be in position
$\p(t) = [x(t),y(t)]^T$ in $\frm{W}$. By denoting with
\[
  \rho_i(t) = \| \p(t) - \an_i \| = \sqrt{(x(t) - x_i)^2 + (y(t) -
    y_i)^2} ,
\]
the actual distance between the tag and the $i$-th anchor, we have
that the ToF $\tof_i(t)$ is given by~\eqref{eq:tof} when
$\rho_{i,m}$ is substituted with $\rho_i(t)$. Therefore, considering that
the $i$-th anchor sends two timestamped messages to the tag at time
$t_i$ and $t_i + \per_i$, respectively, by using~\eqref{eq:AnchorSync}
we get that $\ti^i(t_i) = \ti^m(t_i) + e_{i,m}$ and 
$\ti^i(t_i + \per_i) = \ti^m(t_i + \per_i) + e_{i,m}$). 
These two messages will be received by tag time-scale at 
$\ti(t_i + \tof_i(t_i))$ and $\ti(t_i + \per_i + \tof_i(t_i + \per_i))$. 

Considering the motion of the tag, that can be expressed as 
\[
  \p(t + \per_t) = \p(t) + \per_t \begin{bmatrix}
    v_x(t) \\
    v_y(t)
  \end{bmatrix} = \p(t) + \per_t \ve(t) ,
\]
a first-order discrete time kinematics with a velocity $\ve(t)$, we
have
\[
  \rho_i(t_i + \per_i) = \|\p(t_i) + \per_i \ve(t_i) - \an_i\| ,
\]
which of course depends on the velocity vector $\ve(t_i)$. However, an
upper bound can be found noticing that the maximum increase (or decrease)
of the distance takes place when $\ve(t_i)$ is directed towards the
anchor $\an_i$, i.e., along the direction
$\mathbf{u}_i(t) = [x(t) - x_i, y(t) - y_i]^T$. In such a case,
denoting with $d_i(\per_i) = \|\per_i \ve(t_i)\|$ the tag displacement
taking place at time $t_i$ in the period $\per_i$, we have that
\[
  \rho_i(t_i) - d_i(\per_i) \mathbf{u}_i(t_i) \leq \rho_i(t_i +
  \per_i) \leq \rho_i(t_i) + d_i(\per_i) \mathbf{u}_i(t_i) ,
\]
which yields for the ToF in~\eqref{eq:tof} to
\begin{equation}
  \label{eq:ToFPos}
  \tof_i(t_i) - \frac{d_i(\per_i)}{c} \mathbf{u}_i(t_i) \leq
  \tof_i(t_i + \per_i) \leq \tof_i(t_i) + \frac{d_i(\per_i)}{c}
  \mathbf{u}_i(t_i) .
\end{equation}
For what concerns the TDoA, the last missed ingredients is the
synchronisation algorithm between the tag and the anchor $\an_i$. To
this end, we used an algorithm that is the same of~\eqref{eq:Synchro},
i.e.,
\begin{equation}
  \label{eq:TagSynchro}
  \begin{aligned}
    & \frac{\ti(t_i + \per_{i} + \tof_{i}(t_i + \per_{i})) - \ti(t_i +
      \tof_{i}(t_i))}{\frac{\ti^i(t_i + \per_{i}) - \overline
        o_{i,m}}{\overline \nu_{i,m}} - \frac{\ti^i(t_i) - \overline
        o_{i,m}}{\overline \nu_{i,m}}} = \\
    & = \frac{\ti(t_i + \per_{i} + \tof_{i}(t_i + \per_{i})) - \ti(t_i
      + \tof_{i}(t_i))}{\ti^m(t_i + \per_{i})
      - \ti^m(t_i)} = \\
    & = \frac{\nu (\per_{i} + \alpha \frac{d_i(\per_i)}{c}
      \mathbf{u}_i(t_i))}{\nu_m \per_i} = \overline \nu_{m}, \\
  \end{aligned}
\end{equation}
where
\[
  \overline \nu_{m} = \frac{\nu}{\nu_m} \left (1 + \alpha
    \frac{d_i(\per_i) \mathbf{u}_i(t_i)}{c \per_i} \right ) ,
\]
and $\alpha\in[-1,1]$.

\subsection{Indoor GPS TDoA}

The UTDoA relies on an implicit event: all the anchors receive a tag's
generated broadcast message that acts as an implicit synchronisation event. 
In the case of an indoor GPS-like system with unbounded scalability
-- like the proposed DTDoA --  the messages are transmitted from anchors side to the tags side; 
meaning that a synchronisation event cannot be defined. Strictly speaking, if such a
possibility would exist, the master and anchors would send their packets
simultaneously at time $t_m$. The tag would then measure the difference in reception times as
$\ti(t_m + \tof_m(t_m))$ and $\ti(t_m + \tof_i(t_m))$ and, hence be able to
compute the TDoA as
\begin{equation}
  \label{eq:StandardTDoA}
  \begin{aligned}
    & c\left [\ti(t_m + \tof_i(t_m)) - \ti(t_m + \tof_m(t_m)) \right ]
    = \\
    & = c \nu (\tof_i(t_m) - \tof_m(t_m)) = \nu (\rho_i - \rho_m),
  \end{aligned}
\end{equation}
$\forall i = 1,\dots,n$. Notice that such a measure is only affected
by the relative clock rate $\nu$, which is of course negligible since it
generates an error in the order of some micrometers.

Since such a synchronised event cannot be generated (the anchor will
transmit at slightly different time instants and the tag cannot
receive all the messages at once), the syntonisation between tags and
anchors (i.e., estimation of the relative clock rates) cannot neglected. 
Let's considering a system composed by a master, a number of anchors $i$ 
and a tag. At time $t_m$ a broadcast message is transmitted
from the master, followed by a second message at time
$t_m + \per_m$. The tag timestamps the messages at reception times, denoted as
$\ti(t_m + \tof_m(t_m))$  and $\ti(t_m + \per_m + \tof_m(t_m + \per_m))$ 
Finally the tag stores the two transmission timestamp encapsulated inside the 
broadcast message, defined as $\ti^m(t_m)$ and $\ti^m(t_m + \per_m)$.
The same happens with the anchor $i$ transmitting at time $t_i$ and
$t_i + \per_i$, with tag's timestamped reception times $\ti(t_i + \tof_i(t_i))$
and $\ti(t_i + \per_i + \tof_i(t_i + \per_i))$, and the two transmitted timestamps
$\frac{\ti^i(t_i) - \overline o_{i,m}}{\overline \nu_{i,m}}$
and
$\frac{\ti^i(t_i + \per_{i}) - \overline o_{i,m}}{\overline
  \nu_{i,m}}$, as corrected anchor clock in master time scale
using~\eqref{eq:AnchorSync}. It is then possible to compute at the tag side
the relative clock rates $\overline \nu_{m}$
using~\eqref{eq:TagSynchro} (that explaining the necessity to send/receive two
messages) and a function $g(\per_{i,m})$ of the protocol time interval
$\per_{i,m} = t_i - t_m$ by means of the clock
model~\eqref{eq:TauMaster} as
\begin{equation}
  \label{eq:ProtocolInterval}
  \begin{aligned}
    & g(\per_{i,m}) = \overline \nu_{m} \left (\frac{\ti^i(t_i) -
        \overline o_{i,m}}{\overline \nu_{i,m}} - \ti^m(t_m) \right ) = \\
    & = \overline \nu_{m}
    (\ti^m(t_i) - \ti^m(t_m) + e_{i,m}) = \\
    & = \overline \nu_{m} \nu_m (t_i - t_m) + \overline \nu_{m}
    e_{i,m} = \\
    & = \nu \per_{i,m} + \nu \alpha \frac{d_i(\per_{i,m})
      \mathbf{u}_i(t_m)}{c} + \overline \nu_{m} e_{i,m}.
  \end{aligned}
\end{equation}
By defining the corresponding tag timestamps and the clock
model~\eqref{eq:TauTag} as
\begin{equation}
  \label{eq:TagProtocolInterval}
  \begin{aligned}
    & \ti(t_i + \tof_i(t_i)) - \ti(t_m + \tof_m(t_m)) = \\
    & = \nu (\tof_i(t_i) - \tof_m(t_m)) + \nu (t_i - t_m) = \\
    & = \nu (\tof_i(t_m + \per_{i,m}) - \tof_m(t_m)) + \nu \per_{i,m} ,
  \end{aligned}
\end{equation}
and considering~\eqref{eq:ToFPos} as
\[
  \tof_i(t_m + \per_{i,m}) = \tof_i(t_m) + \alpha
  \frac{d_i(\per_{i,m})}{c} \mathbf{u}_i(t_m) ,
\]
we can finally have derive the DTDoA equation: 
\begin{equation}
  \label{eq:GPSTDoA}
  \begin{aligned}
    & c \left [\ti(t_i + \tof_i(t_i)) - \ti(t_m + \tof_m(t_m)) -
      g(\per_{i,m})\right ] = \\
    & = \nu (\rho_i - \rho_m) - c \overline \nu_{m} e_{i,m}.
  \end{aligned}  
\end{equation}
computed using~\eqref{eq:tof} and~\eqref{eq:ProtocolInterval}.

Those quantities can be equivalently computed for the delayed
messages, by considering $\ti^m(t_m + \per_m)$ and
$\ti^i(t_i + \per_i)$ in~\eqref{eq:ProtocolInterval}; and
$\ti(t_m + \per_m + \tof_m(t_m + \per_m))$ and
$\ti(t_i + \per_i + \tof_i(t_i + \per_i))$
in~\eqref{eq:TagProtocolInterval}. Of course,
comparing~\eqref{eq:GPSTDoA} and~\eqref{eq:StandardTDoA}, we can
notice that in the presence of multiple time sources 
(i.e., having $i$ anchors) induces potential errors
stemming from the imperfect synchronisation with the master,
highlighted in~\eqref{eq:AnchorSync}, which is, unfortunately unavoidable.

\section{Uncertainties}
\label{sec:uncertainties}

The main source of uncertainties -- neglecting the effects of 
ageing or the drift changes induced by harsh environmental conditions
(e.g., mechanical vibrations or temperature
effects~\cite{FontanelliMWOS11}) -- is related to timestamping accuracy
operation. Albeit those effects can be dramatically reduced by
implementing double consecutive message transmissions (i.e., the
timestamp is acquired when the message is sent, and then transmitted
in a second message, as in~\cite{FontanelliMWOS11}), the effect cannot
be entirely removed.  As a consequence, by denoting with $\ts$ the
generic timestamping uncertainty affecting each clock in the network,
we can rewrite the clock models as
\begin{equation}
  \label{eq:TauUnc}
  \begin{aligned}
    \m{\ti^m}(t) & = o_m + \nu_m t + \ts_m(t) = \ti^m(t) + \ts_m(t), \\
    \m{\ti^i}(t) & = o_i + \nu_i t + \ts_i(t) = \ti^i(t) + \ts_i(t), \\
    \m{\ti}(t) & = o + \nu t + \ts(t) = \ti(t) + \ts(t) ,
  \end{aligned}
\end{equation}
where we use the superscript $\m{\cdot}$ to denote each measurement
result. It worth to be noted that we implicitly assume that the measurement
uncertainties $\ts_m(t)$, $\ts_i(t)$ and $\ts(t)$ are random variables
generated by white, stationary and zero mean processes with variances
$\sigma_{\ts_m}^2$, $\sigma_{\ts_i}^2$ and $\sigma_{\ts}^2$,
respectively.

By assuming a first-order Taylor approximation for the uncertainty, the
relative clock rates $\overline \nu_{i,m}$ in~\eqref{eq:Synchro}, when
computed with~\eqref{eq:TauUnc} becomes
\[
  \frac{\m{\ti^i}(\overline t + \per_{i,m} + \tof_{i,m}) -
    \m{\ti^i}(\overline t + \tof_{i,m})}{\m{\ti^m}(\overline t +
    \per_{i,m}) - \m{\ti^m}(\overline t)} \approx \overline \nu_{i,m}
  + \beta = \m{\overline \nu_{i,m}},
\]
having the uncertainty mean $\mu_\beta = \E{\beta} = 0$ (the
$\E{\cdot}$ is the usual expected operator) and its variance
\[
  \E{\beta^2} = \sigma_\beta^2 = \frac{2}{\nu_m^2 \per_{i,m}^2}
  (\sigma_{\ts_i}^2 + \overline \nu_{i,m}^2 \sigma_{\ts_m}^2) ,
\]
It can thus be argued that the larger the interval $\per_{i,m}$, 
the smaller the uncertainty on the indirect measurement of $\m{\overline \nu_{i,m}}$. 
For the relative offsets $\overline o_{i,m}$ in~\eqref{eq:SynchroOffset},
we have similarly
\begin{equation}
  \label{eq:Gamma}
  \m{\ti^i}(\overline t + \tof_{i,m}) - \m{\overline \nu_{i,m}}
  \m{\ti^m}(\overline t) - \tof_{i,m} \approx \overline o_{i,m} +
  \gamma(\overline t) = \m{\overline o_{i,m}}(\overline t),
\end{equation}
with mean $\mu_{\gamma}(\overline t) = \E{\gamma(\overline t)} = 0$
and variance
\[
  \begin{aligned}
    & \sigma_{\gamma}^2(\overline t) = \E{\gamma(\overline t)^2} =
    \left ( 1 - 2 \frac{\tau^m(\overline
        t)}{\nu_m \per_{i,m}} \right ) \sigma_{\ts_i}^2 + \\
    & + \overline \nu_{i,m}^2 \left ( 1 + 2 \frac{\tau^m(\overline
        t)}{\nu_m \per_{i,m}} \right ) \sigma_{\ts_m}^2 +
    \tau^m(\overline t)^2 \sigma_\beta^2 .
  \end{aligned}
\]
By using~\eqref{eq:SynchroOffset}, we can also
make use of the quantities delayed by $\per_{i,m}$ yielding to a
similar quantity as in~\eqref{eq:Gamma} but denoted with
$\m{\overline o_{i,m}}(\overline t + \per_{i,m})$. 
It is then possible to formulate a new estimate as
\begin{equation}
  \label{eq:AverageRelOff}
  \hat{\overline o}_{i,m} = \frac{\m{\overline o_{i,m}}(\overline t) +
    \m{\overline o_{i,m}}(\overline t + \per_{i,m})}{2} = \overline
  o_{i,m} + \overline{\gamma}(\overline t),
\end{equation}
that is again affected by a zero mean uncertainty with variance
\[
  \sigma_{\overline{\gamma}}^2(\overline t) =
  \frac{\sigma_{\ts_i}^2}{2} + \frac{\sigma_{\ts_m}^2}{2} +
  \tau^m\left (\overline t + \frac{\per_{i,m}}{2} \right )^2
  \sigma_\beta^2 ,
\]
which may or may not be more useful than~\eqref{eq:Gamma} depending on
the value of $\per_{i,m}$. This is a direct consequence of the fact
that $\m{\overline o_{i,m}}(\overline t)$ and
$\m{\overline o_{i,m}}(\overline t + \per_{i,m})$ are correlated by
$\m{\overline \nu_{i,m}}$.

For what concerns equation~\eqref{eq:AnchorSync}, if the time $t$ in which the
anchor time scale is converted into the master time scale is different
from $\overline t$, $\overline t + \per_{i,m}$,
$\overline t + \tof_{i,m}$ and $\overline t + \per_{i,m}+ \tof_{i,m}$
(i.e., the times in which~\eqref{eq:Synchro}
and~\eqref{eq:SynchroOffset} are used), we can again use a first-order
Taylor approximation to re-write~\eqref{eq:AnchorSync} as
\[
  \frac{\m{\ti^i}(t) - \hat{\overline o}_{i,m}}{\m{\overline
      \nu_{i,m}}} \approx \ti^m(t) + e_{i,m} + \varepsilon(t) .
\]
Having $\E{\varepsilon(t)} = 0$ and
\[
  \begin{aligned}
    \sigma_{\varepsilon}^2(t) & = \frac{\sigma_{\ts_i}^2}{\overline
      \nu_{i,m}^2} + \frac{\sigma_{\overline{\gamma}}^2(\overline
      t)}{\overline \nu_{i,m}^2} + \\
    & + \left [ \frac{(\tau^m(t) + e_{i,m})^2}{\overline \nu_{i,m}^2}
      - 2 \frac{\tau^m(t) + e_{i,m}}{\overline \nu_{i,m}^2}
      \tau^m(\overline t) \right ] \sigma_\beta^2 .
  \end{aligned}
\]
allow us to compute the uncertain version of~\eqref{eq:TagSynchro} as
\[
  \begin{aligned}
    & \frac{\m{\ti}(t_i + \per_{i} + \tof_{i}(t_i + \per_{i})) -
      \m{\ti}(t_i + \tof_{i}(t_i))}{\frac{\m{\ti^i}(t_i + \per_{i}) -
        \hat{\overline o}_{i,m}}{\m{\overline \nu_{i,m}}} -
      \frac{\m{\ti^i}(t_i) - \hat{\overline
          o}_{i,m}}{\m{\overline \nu_{i,m}}}} \approx \\
    & \approx \overline \nu_{m} + \xi(\per_i) = \m{\overline \nu_{m}} , \\
  \end{aligned}
\]
with $\E{\xi(\per_i)} = 0$ and
\[
  \sigma_{\xi}^2(\per_i) = \frac{2 \overline \nu_{m}^2}{\nu_i^2
    \per_i^2} \sigma_{\ts_i}^2 + \frac{2}{\nu_m^2 \per_i^2}
  \sigma_{\ts}^2 + \frac{\overline \nu_{m}^2}{\nu_{i,m}^2}
  \sigma_{\beta}^2 .
\]
Similarly, to compute the noisy protocol interval
in~\eqref{eq:ProtocolInterval}, we have
\[
  \begin{aligned}
    \m{g}(\per_{i,m}) & = \m{\overline \nu_{m}} \left (\frac{\m{\ti^i}(t_i) - \hat{\overline o}_{i,m}}{\m{\overline \nu_{i,m}}} - \m{\ti^m}(t_m) \right ) \approx \\
    & \approx g(\per_{i,m}) + \varphi(t_i, t_m) .
  \end{aligned}
\]
In this case $\E{\varphi(t_i, t_m)} = 0$ and
\[
  \begin{aligned}
    & \sigma_{\varphi}^2(t_i, t_m) = \overline \nu_m^2
    (\sigma_{\varepsilon}^2(t_i) + \sigma_{\nu_m}^2) + \\
    & + \frac{2 \overline \nu_m^2}{\nu_m \per_i} (\nu_m \per_{i,m} +
    e_{i,m}) (\sigma_{\varepsilon}^2(t_i) - \sigma_{\varepsilon}(t_i,
    t_i + \per_i)) + \\
    & + (\nu_m \per_{i,m} + e_{i,m})^2 \sigma_{\xi}^2(\per_i) ,
  \end{aligned}
\]
where $\sigma_{\varepsilon}(t_i, t_i + \per_i))$ is the correlation
between the uncertainties $\varepsilon(t_i)$ and
$\varepsilon(t_i + \per_i)$ leading to
\[
  \begin{aligned}
    & \E{\varepsilon(t_i) \varepsilon(t_i + \per_i)} =
    \frac{(\tau^m(t_i) + e_{i,m})(\tau^m(t_i +
      \per_i)}{\overline \nu_{i,m}^2} \sigma_\beta^2 + \\
    & + \frac{\sigma_{\overline{\gamma}}^2(\overline t)}{\overline
      \nu_{i,m}^2} - \frac{\tau^m(t_i) + \tau^m(t_i + \per_i) +
      2e_{i,m}}{\overline \nu_{i,m}^2} \tau^m\left (\overline t +
      \frac{\per_{i,m}}{2} \right ) \sigma_\beta^2 .
  \end{aligned}
\]

We are now ready to conclude the uncertainty analysis by computing the
final DTDoA relation~\eqref{eq:GPSTDoA} once the measured quantities
in~\eqref{eq:TauUnc} are used:
\[
  \begin{aligned}
    & c \left [\m{\ti}(t_i + \tof_i(t_i)) - \m{\ti}(t_m + \tof_m(t_m))
      -
      \m{g}(\per_{i,m})\right ] = \\
    & = \nu (\rho_i - \rho_m) - c \overline \nu_{m} e_{i,m} +
    \lambda(t_i, t_m) ,
  \end{aligned}
\]
resulting in an overall uncertainty with mean
$\E{\lambda(t_i, t_m)} = 0$ and variance
\begin{equation}
  \label{eq:FinalTDoAUnc}
  \sigma_{\lambda}^2(t_i, t_m) = c^2 \left [ 2 \left (1 + \frac{\nu_m
        \per_{i,m} + e_{i,m}}{\nu_m \per_i} \right ) \sigma_{\ts}^2 +
    \sigma_{\varphi}^2(t_i, t_m) \right ] .
\end{equation}

\subsection{Numerical analysis}

\begin{figure}[t]
  \centering
  \includegraphics[width=0.9\columnwidth]{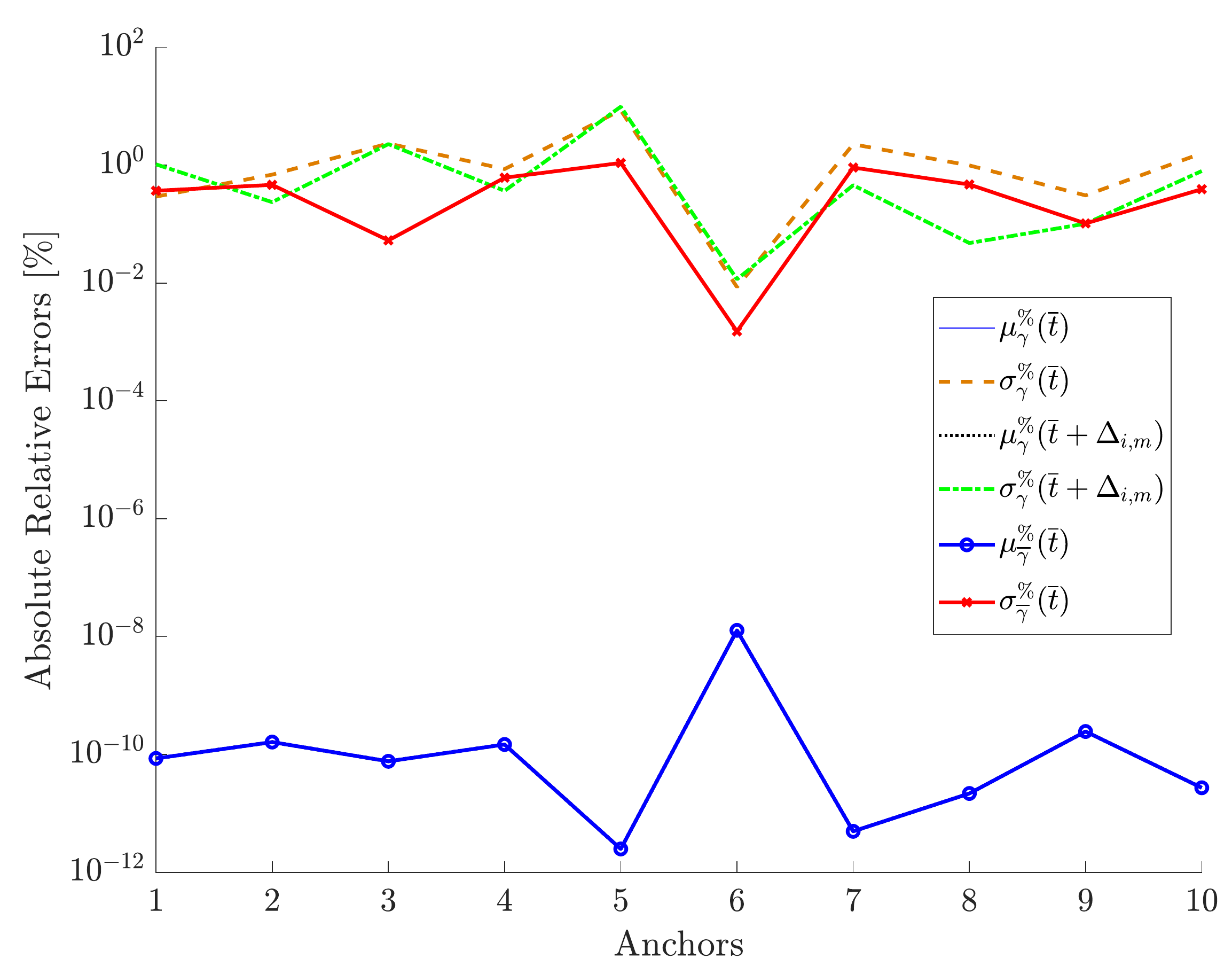}
  \caption{Absolute relative error of the theoretical approximations
    with respect to Monte Carlo simulations for the random variables
    $\gamma(\overline t)$, $\gamma(\overline t + \per_{i,m})$ and
    $\overline{\gamma}(\overline t)$.}
  \label{fig:GammaApprox}
  \vspace{-1em}
\end{figure}

In this section we present the simulations results about the validity 
of the uncertainty analysis presented in the previous subsection.
The first simulation analyse the tightness of the approximations of the 
uncertainties introduced by the first-order Taylor linearisation. 
Figure~\ref{fig:GammaApprox} presents the absolute relative 
error, expressed in percentage, between the theoretical values of the 
mean $\mu_{\gamma}(\overline t)$ and the standard deviation
$\sigma_{\gamma}(\overline t)$ in~\eqref{eq:Gamma} for the uncertainty
$\gamma(\overline t)$, with respect to the values obtained through
$10000$ Monte Carlo Simulations, here denoted with
$\mu_{\gamma}^{MC}(\overline t)$ and $\sigma_{\gamma}^{MC}(\overline t)$:
\[
  \mu_{\gamma}^\%(\overline t) = 100 \frac{|\mu_{\gamma}(\overline t) -
    \mu_{\gamma}^{MC}(\overline t)|}{\mu_{\gamma}^{MC}(\overline t)} ,
\]
and
\[
  \sigma_{\gamma}^\%(\overline t) = 100
  \frac{|\sigma_{\gamma}(\overline t) - \sigma_{\gamma}^{MC}(\overline
    t)|}{\sigma_{\gamma}^{MC}(\overline t)} ,
\]
respectively.
%-%
%-%
The Monte Carlo simulations assume $10$ anchors (on the abscissa of
Figure~\ref{fig:GammaApprox}) with randomly generated clock time
offsets, clock rates, positions, delays $\per_i$ and $\per_{i,m}$, having
zero-mean, Gaussian time stamping uncertainties $\ts_m(t)$, $\ts_i(t)$,
and $\ts(t)$, with a maximum value of plus/minus $15.65$~ps (i.e., one
clock tick, as per the available hardware, see
Section~\ref{sec:ExperimentalResults}). As can be seen from
Figure~\ref{fig:GammaApprox}, whatever uncertainty is considered, i.e.
$\gamma(\overline t)$, $\gamma(\overline t + \per_{i,m})$ or the
average of the two $\overline{\gamma}(\overline t)$ reported
in~\eqref{eq:AverageRelOff}, the theoretical analysis matches
remarkably well with the simulations (the mean, for example, are all
overlapped with negligible errors in
Figure~\ref{fig:GammaApprox}). Similar results are obtained for all
the other quantities, here not reported for space limits.

\begin{figure}[t]
  \centering
  \begin{tabular}{cc}
    \includegraphics[width=0.45\columnwidth]{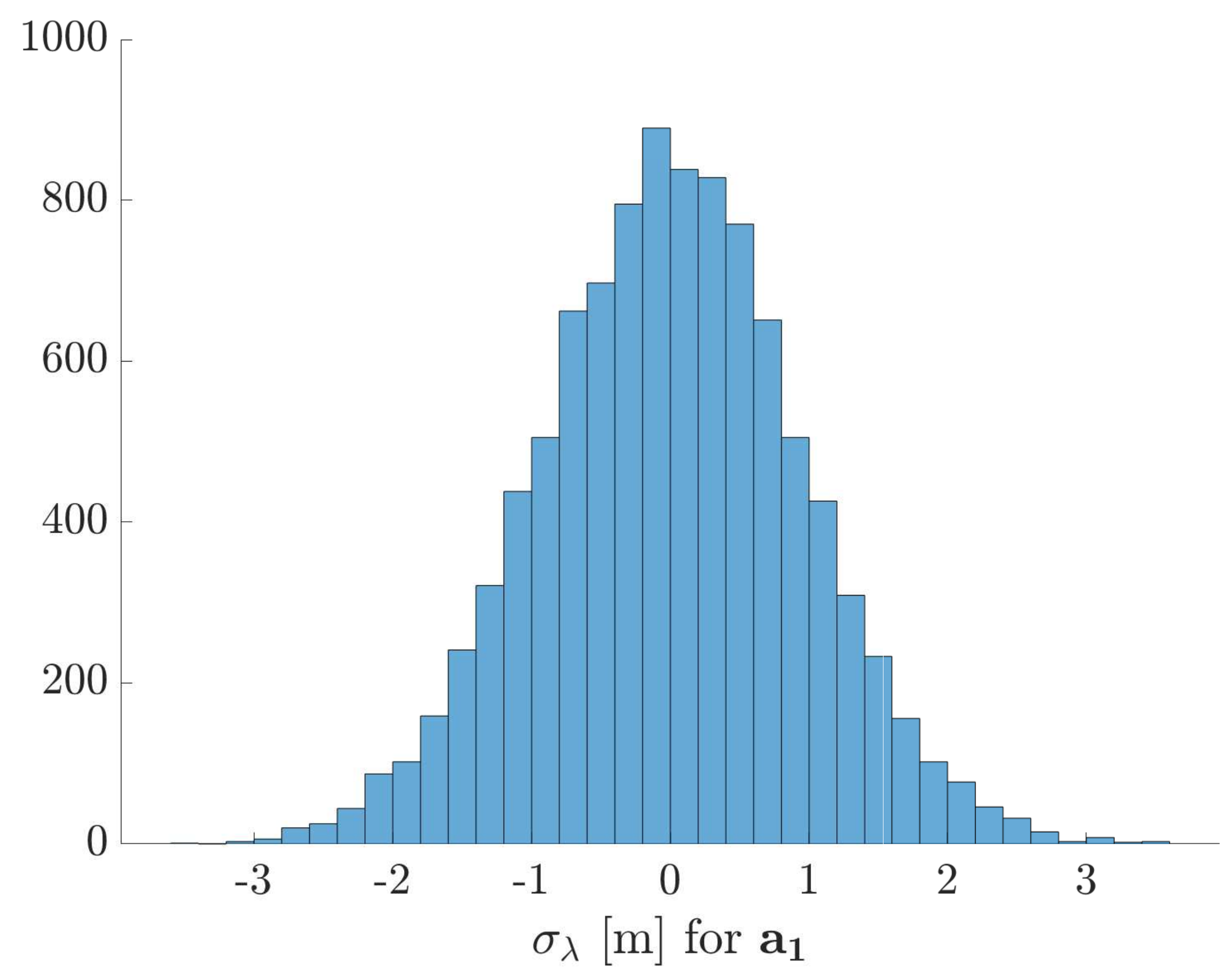} &
                                                                 \includegraphics[width=0.45\columnwidth]{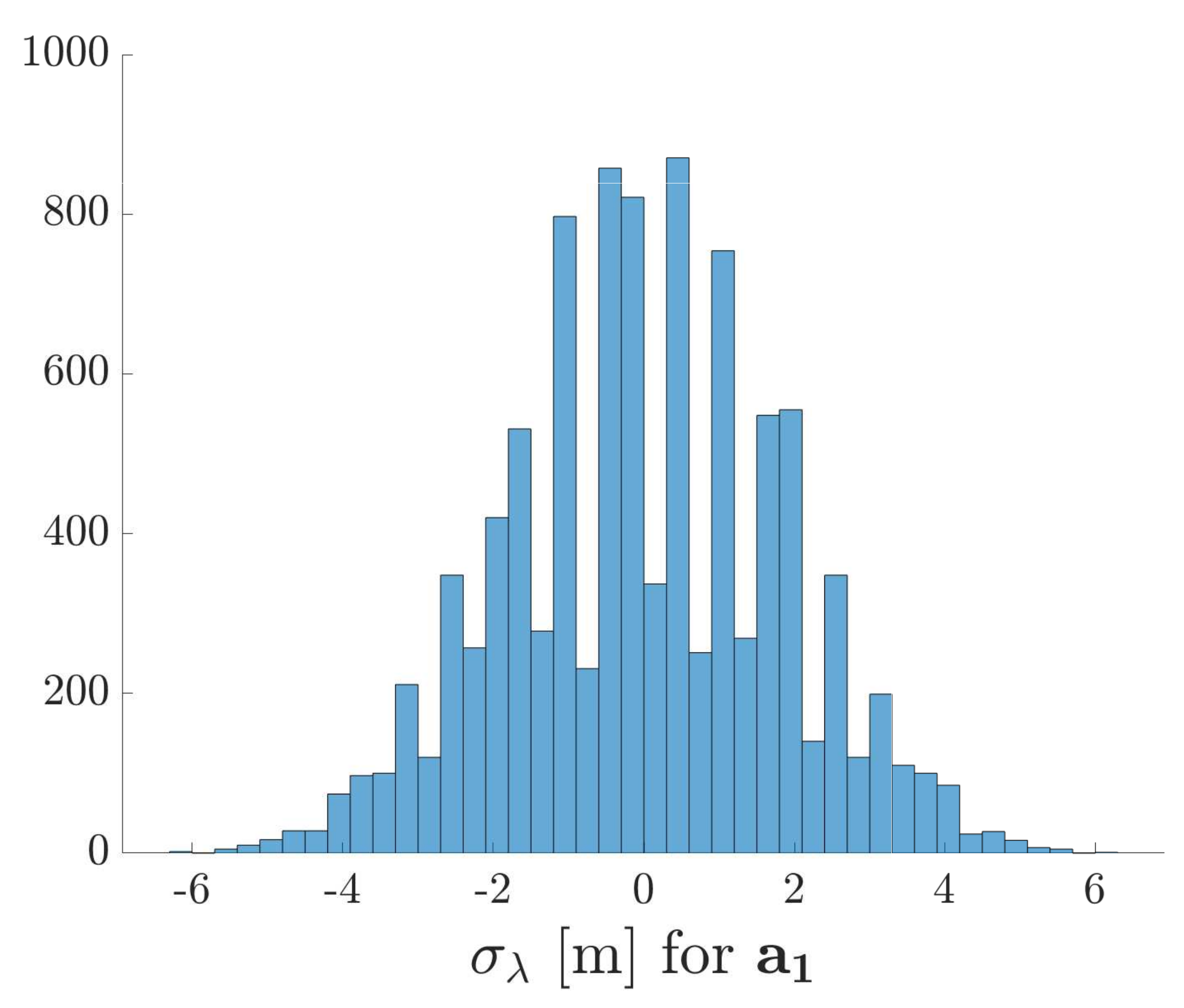}
    \\
    (a) & (b)
  \end{tabular}
  \caption{Distribution of the DTDoA uncertainties
    $\sigma_{\lambda}^2(t_i, t_m)$ in~\eqref{eq:FinalTDoAUnc} in the
    case of (a) Gaussian and (b) Uniform timestamping uncertainties.}
  \label{fig:TDoAApprox}
  \vspace{-1em}
\end{figure}

Of course, the approximation becomes less accurate when the
timestamping uncertainties are uniformly distributed between
plus/minus $15.65$~ps, as reported in Figure~\ref{fig:TDoAApprox} 
that presents the DTDoA uncertainties in~\eqref{eq:FinalTDoAUnc} 
for the anchor $\an_1$.
%-%
%-%

\section{Experimental Results}
\label{sec:ExperimentalResults}

To evaluate and validate the statistical model discussed in the
previous section, we have implemented the proposed DTDoA algorithm
using UWB radios. UWB radio transmission is based upon impulse radio
(IR) schemes. The pulses being applied in IR have a very short
temporal duration (typically in order of nanoseconds), which results
in an ultra-wide band spectrum~\cite{oppermann2004uwb}. This very large
spectrum allows to achieve excellent time and spatial resolution by
evaluating the Channel Impulse Response (CIR) of the signal.

\subsection{Hardware Implementation}

To implement our LPS, we decided to use the commercial-off-the-shelf
(COTS) Decawave
DWM1001\footnote{https://www.decawave.com/sites/default/files/dwm1001\_datasheet.pdf}
SoM, which attracted many interests for implementing indoor
positioning systems~\cite{7962828}. The DWM1001 is a compact module
that integrates both a low power nRF52832 MCU and the Decawave
DW1000\footnote{https://www.decawave.com/sites/default/files/resources/dw1000-datasheet-v2.09.pdf}
UWB transceiver. It also integrates RF circuitry, an UWB antenna and a
motion sensor for sensor fusion
applications~\cite{Frisk_Nilsson_2014}.

The DW1000 chip is an IEEE 802.15.4-2011~\cite{ieee2011ieee} compliant
UWB transceiver which can operate on 6 different frequency bands with
centre frequencies between 3.5 to 6.5 GHz and a bandwidth of 500 or
900 MHz. It provides the possibility of ranging measurements and
retrieving the measured CIR. The chip also offers three different data
rates: 110~kbps, 850~kbps and 6.8~Mbps.

The DW1000 clocking scheme is based on 3 main circuits; crystal
oscillator (trimmed in production to reduce the initial frequency
error to approximately 3 ppm), Clock Phase-Locked Loop (PLL) and RF
PLL. The on-chip oscillator is designed to operate at a frequency of
38.4~MHz. This clock is then used as the reference input to the two
on-chip PLLs. The clock PLL generates a 63.8976~GHz reference clock
required by the digital backend for signal processing. The RF PLL
generates the clock for the receive and transmit chain. The DW1000
automatically timestamps transmitted and received frames with a
precision of 40-bit. In turns, working at a nominal 64~GHz resolution,
packets are timestamped with a 15.65~ps event timing
precision\footnote{https://www.decawave.com/dw1000/usermanual/}.

During the experimental tests, the DWM1001 SoM was configured to use
UWB Channel 5 ($f_c=6489.6$~MHz, $BW=499.2$~MHz), a preamble length of
128 symbols, the highest Pulse Rate $PR=64$~MHz, and the highest Data
Rate$DR=6.8$~Mbps.

\subsection{Network infrastructure}

To exploit the UWB protocol to implement an LPS, a specific number of
DWM1001 module must be programmed to act as anchors and, hence,
provide a reference infrastructure for the tags. In our case, the
reference infrastructure comprises six reference anchors mounted on
the wall of our laboratory. Each anchor is mainly composed of a
Raspberry PI 3 and a DWM1001 module. Finally, data sharing and
acquisition is implemented by leveraging the MQTT protocol. In this
way, both anchors and tags data can be retrieved easily also by a
remote system.

\subsection{Positioning Results}

\begin{figure}[t]
  \centering
  \begin{tabular}{c}
    \includegraphics[width=1\columnwidth]{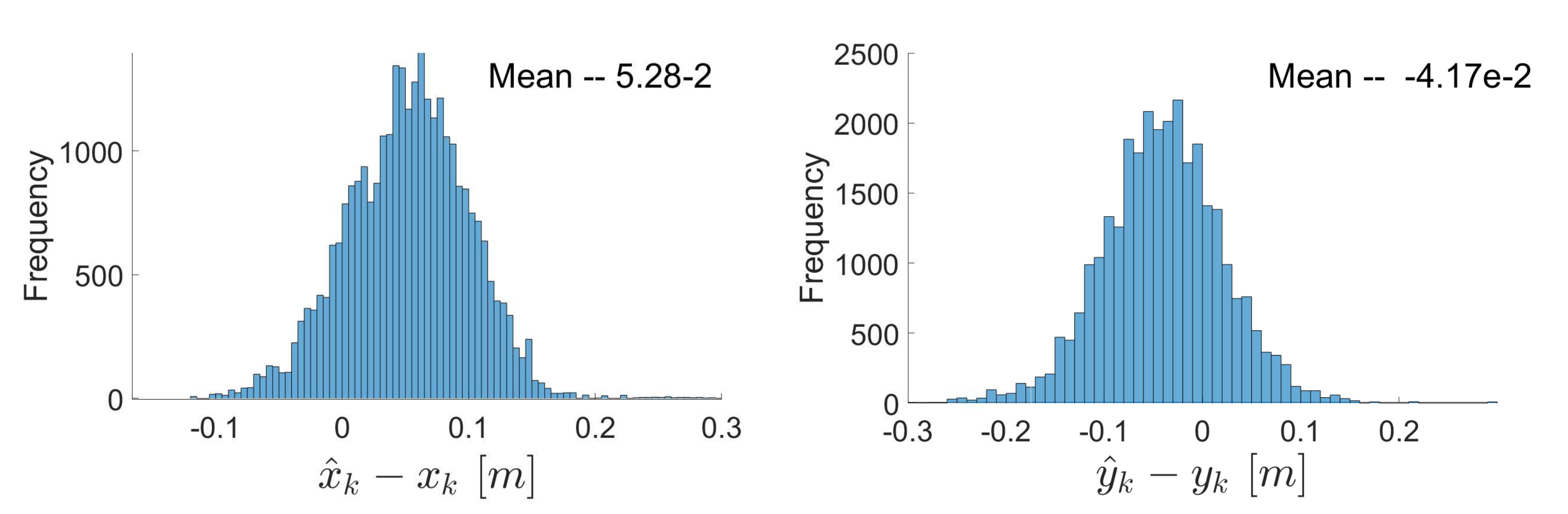} \\
    Test position 1 \\
    \includegraphics[width=1\columnwidth]{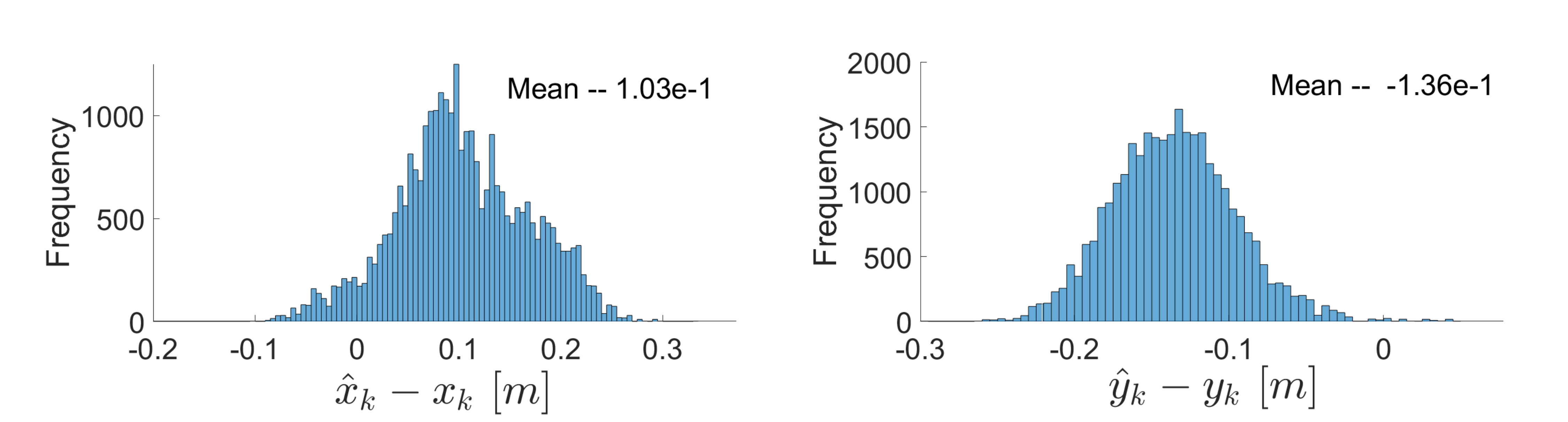} \\
    Test position 2
  \end{tabular}
  \caption{Histogram of the error for the test positions 1 and 2 along
    the $X_w$ (left) and $Y_w$ (right) reference axes over $30000$
    repeated measurements.}
  \label{fig:HistogramErrorXY}
  %\vspace{-1em}
\end{figure}

To evaluate and characterise the statistical properties of the
developed positioning infrastructure, two experiments were
conducted. The first test is done in static conditions with a tag
positioned in two known locations. Figure~\ref{fig:HistogramErrorXY}
depicts the histogram of the estimation error along the $X_w$ and
$Y_w$ axis, i.e.  $\tilde x_k = \hat x_k - x_k$ and
$\tilde y_k = \hat y_k - y_k$, respectively, for the two test
positions over $30000$ repeated measurements.
%-%
%-%
%-%
\begin{figure}[t]
	\centering
	\includegraphics[width=0.7\columnwidth]{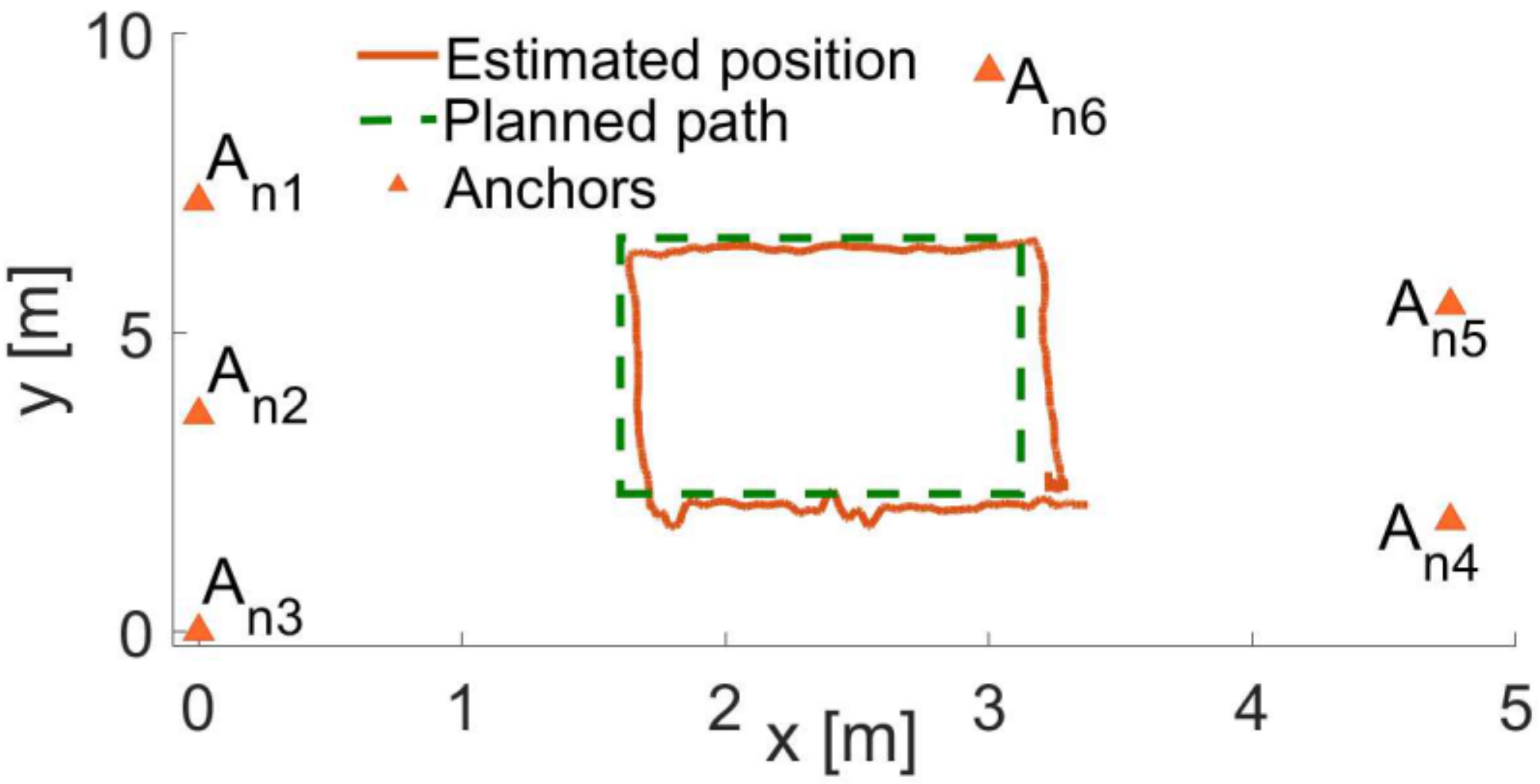}
	\caption{Walking test}
	\label{fig:Walk}
	\vspace{-1em}
\end{figure}
%-%
As can be noted, they present a non-zero mean Gaussian distribution
due to the effects of angle-dependent UWB pulse distortion and the
path overlap~\cite{UWBbias}. Moreover, the measure is also subjected
to the anchors distribution in the testing room that may cause
reflections due to the presence of metallic objects. Nonetheless, the
precision is below the claimed $20$~cm of error.

\begin{figure}[t]
	\centering
	\includegraphics[width=0.7\columnwidth]{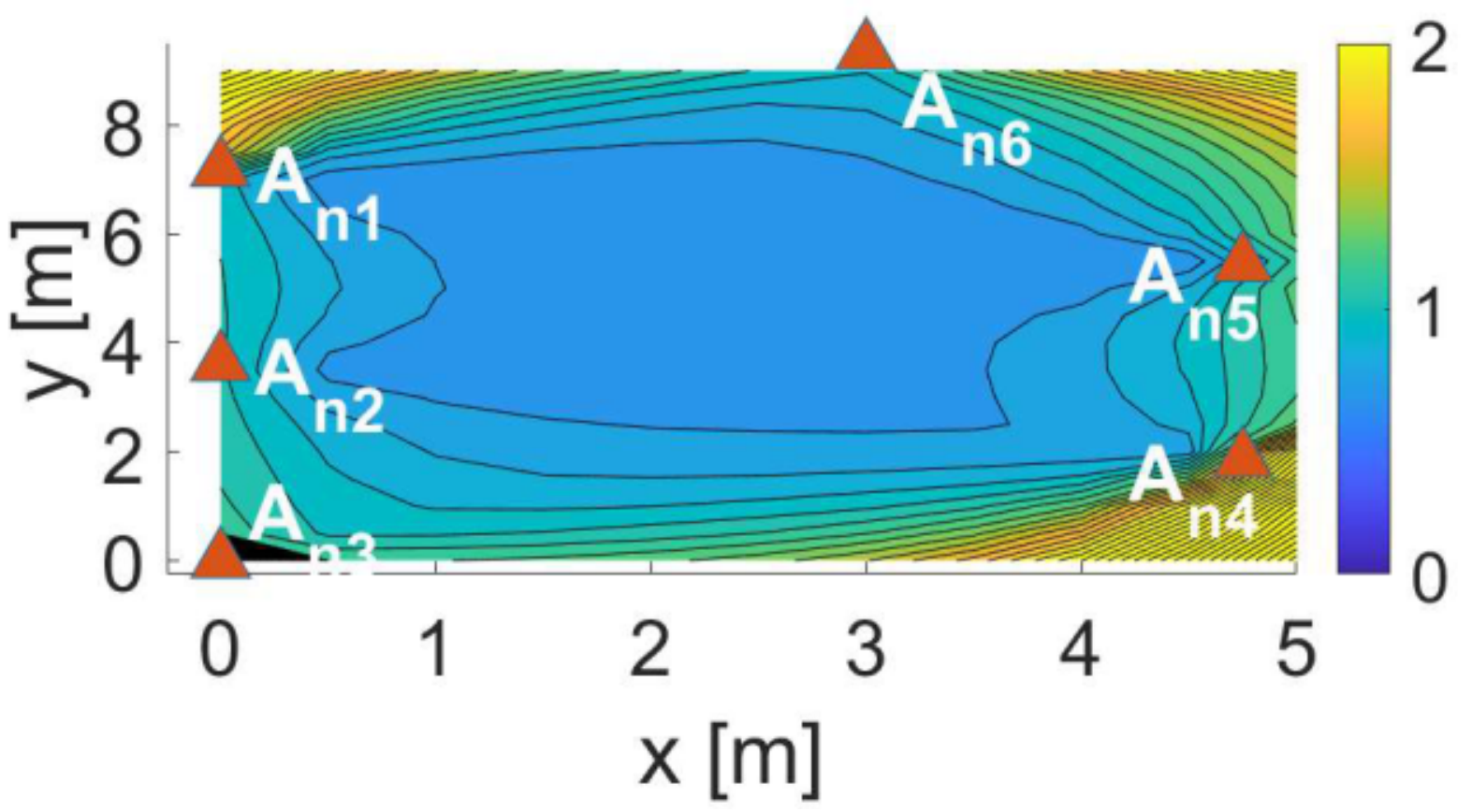}
	\caption{PDoP map for the actual anchors distributions for the
          experiment in Figure~\ref{fig:Walk}.}
	\label{fig:Pdopmap}
	\vspace{-1em}
\end{figure}

The second test was conducted while moving the receiver along a
planned path to evaluate the effectiveness of the infrastructure, both
for tracking or navigation purposes. The result is reported in
Figure~\ref{fig:Walk} where the green dashed line represents the
planned path.  The asymmetry in the estimated positions, can be
explained using the Position Dilution of Precision (PDoP) (see
Figure~\ref{fig:Pdopmap} for the PDoP map of the adopted environment).
%-%
%-%
Although the PDoP does not consider complex non-line-of-sight and
multipath problems, it can be considered a valid simple metric to
express positioning accuracy~\cite{DOP}.

\begin{figure}[t]
  \centering
  \includegraphics[width=0.7\columnwidth]{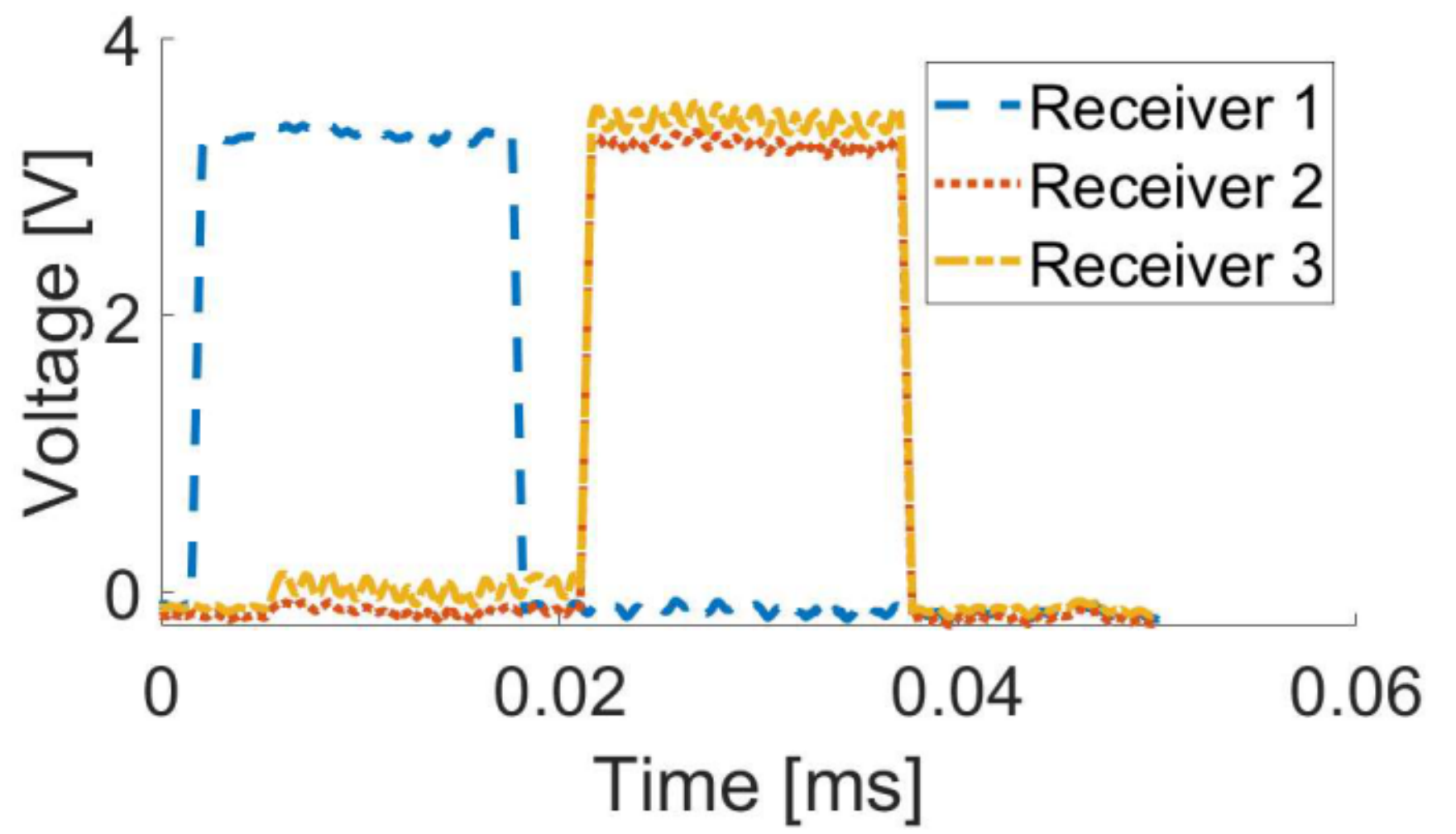}
  \caption{Three tags complete independently their position estimation
    phase. The graph reports the voltage measured from a pin of the
    GPIO, which is rose when a tag is engaged into the positioning
    phase.}
  \label{fig:Picoscope}
  \vspace{-1em}
\end{figure}

As described in Section~\ref{sec:Intro}, the main purpose of the
presented solution is to determine the position of an unlimited number
of assets (e.g., robots) inside the environment. As presented in
Figure~\ref{fig:Picoscope}, which shows the execution time needed for
receiving all the messages from the reference anchors, the positioning
routine is scheduled at the receiver side.
%-%
%-%
In this example, Receiver 1 completes its pose estimation before the
others, while the second and the third receivers estimate their
position simultaneously. This confirms that the system allows
simultaneous position estimates from different tags, thus empirically
prove that the system can scale up to infinity.

\section{Conclusion}
\label{sec:conclusion}

This paper presents the solution to the scalability problem of
LPSs. The introduced DTDoA scheme enables the possibility to supply
the pose information to an infinite number of entities and robots
endowed with an UWB receiver. The source of uncertainties analysis
highlighted the critical points for the DTDoA scheme and guided the
development of an LPS with a maximum absolute position error below
$20$~cm. The maximum pose information update rate reached in this work
is $67$~Hz. In future work, we investigate alternative downlink
schemes to increase the update rate of the localisation, removing the
limit given by the number of active anchors. Another goal is to
increase the accuracy of localisation system exploiting hybrid
techniques to take care of the non-line-of-sight and multipath
problems.

\section*{Acknowledgement}
This work was supported by the Italian Ministry for Education, University and Research (MIUR) under the program “Dipartimenti di Eccellenza (2018-2022)”. 

%\newpage

\bibliographystyle{IEEEtran}
\bibliography{refs}

\end{document}